\documentclass[aps,prl,preprint,groupedaddress]{revtex4-1}
\usepackage{graphicx}
\usepackage{amsmath}

\begin{document}

\title{Local chirality of optical waves in ultrasmall resonators}

\author{Brandon Redding$^1$, Li Ge $^2$, Qinghai Song$^3$, Jan Wiersig $^4$, Glenn S. Solomon $^5$, Hui Cao $^1$}

\affiliation{$^1$ Department of Applied Physics, Yale University, New Haven, CT 06520, USA\\
$^2$ Department of Electrical Engineering, Princeton University, Princeton, New Jersey 08544, USA \\
$^3$ National Key Laboratory of Tunable Laser Technology, Institute of Opto-Electronics, Harbin Institute of Technology, Harbin, 150080, China \\
$^4$ Institut f{\"u}r Theoretische Physik, Universit{\"a}t Magdeburg, Postfach 4120, D-39016 Magdeburg, Germany\\
$^5$ Joint Quantum Institute, NIST and University of Maryland, Gaithersburg, Maryland 20899, USA}

\date{\today}

\begin{abstract}

The local chiral symmetry between clockwise (CW) and counter-clockwise (CCW) propagating light in a deformed microcavity can be broken by wave optics effects, which become significant as the cavity size approaches the wavelength. 
We show that the spatial separation of the CW and CCW ray orbits underlying the high quality factor resonant modes results in unidirectional emission in free space.  
In the presence of a waveguide, evanescent coupling also becomes directional, and the output direction can be varied by selecting the coupling position along the cavity boundary. 
Our results demonstrate that the local chirality can be utilized to control the output directionality and enhance the collection efficiency of emission from ultrasmall resonators. 
 
\end{abstract}

\maketitle

Chirality has important implications in many areas of physics including optics. 
Typically, chirality is introduced via structural geometry with distinct left- and right-handed
forms.
In this Letter, we will use optical microdisk resonators as an example to illustrate that even if the structure is not chiral, the {\it local} symmetry between clockwise (CW) and counterclockwise (CCW) propagating waves may still be broken.
A dielectric microdisk can confine light by total internal reflection at the boundary enabling high quality factors and lasing \cite{McCallAPL92}.
If the microdisk is perfectly circular, the rotational symmetry of the cavity shape leads to isotropic output, which hinders light collection. 
One approach to breaking the rotational symmetry is positioning a waveguide (or fiber) sufficiently close to the cavity to allow the emission to couple to the waveguide evanescently \cite{kippenberg_PRA06, painter_Nature07, choi_PTL03}.
The chiral symmetry between the CW and CCW waves propagating in the cavity is preserved and the energy coupled out is split evenly into the two waveguide directions. 
Unidirectional coupling to a waveguide can be achieved by introducing chirality to the cavity shape,  e.g. a spiral-shaped cavity \cite{chern_APL03, poon_OE07, wiersig_OE08, poon_OE08}; however, this approach results in a significant reduction in the quality factor ($Q = \omega / \Delta \omega$, where $\omega$ is the resonance frequency and $\Delta \omega$ is the linewidth). 

An approach to achieve non-isotropic emission without introducing a waveguide is to deform the cavity from a circular shape \cite{nockel_OL94, stone_nature97, gmachl_Science98, harayama_PRL03, lebental_APL06, ryu_PRE06, lee_PRA07, gao_APL07, wiersig_PRL08, song_PRA09, yan_APL09,shinohara_PRA09, yi_APL09}.
In most cases studied so far, the deformation does not introduce chirality to the cavity shape.  Nonetheless, directional emission is possible when the rotational symmetry is removed.
When the disk radius $R$ is much larger than the optical wavelength $\lambda$ (in vacuum), intracavity ray dynamics determine the output directionality, and can be manipulated via deliberate deformation of the cavity shape \cite{schwefel_JOSAB04} to produce emission in a single direction \cite{wiersig_PRL08}. 
Recently, wavelength-scale deformed cavities were fabricated in efforts to reduce the cavity mode volume \cite{song_PRL10}.
As $kR$ ($k = 2 \pi / \lambda$) approaches one, the classical ray model breaks down, and wave optical phenomena become dominant. 
High-$Q$ modes may be formed by partial barriers in the phase space \cite{shim_PRA11}, and their emission (to free space) is not as directional as from larger cavities. 
The only method that has been shown to generate unidirectional output from such small cavities is accidental coupling of an isotropic high-$Q$ mode (HQM) to an anisotropic low-$Q$ mode (LQM) \cite{song_PRL10}.
In this Letter, we will show that when we move further into the wave regime by making the cavity even smaller, the HQMs become directional again, in the absence of mode coupling. 
Such behavior is attributed to the breaking of the local balance between CW and CCW wave amplitudes. 
Since the geometric shape of the cavities in our study is nonchiral, the global chiral symmetry is preserved for each optical resonance; namely, the total amounts of CW and CCW waves are equal. 
Only locally is the chiral symmetry broken by wave optics effects. 
Such local symmetry breaking also enables unidirectional coupling to the waveguide, and the output direction can be varied by selecting the coupling point on the cavity boundary.  
Experimentally, we realized lasing at $kR$ as small as 3, and observed unidirectional emission from the lasing modes. 
In addition, we demonstrated selective coupling of the CW or CCW waves in a lasing mode to a waveguide placed tangential to the disk boundary.      
The combination of relatively high-$Q$, small mode volume, and in-plane directional emission or directional waveguide coupling are appealing not only to the fundamental studies of cavity quantum electrodynamics, but also to the developments of nanophotonic devices such as ultrasmall light sources, optical switches and sensors. 

Though our results are relevant for all deformed cavities deep in the wave optics regime, we focused on the lima\c{c}on cavity whose boundary is described by $r(\theta) = R(1+ \epsilon \cos \theta)$ in polar coordinates. 
The cavity shape is nonchiral, $r(-\theta) = r(\theta)$. 
The refractive index of the cavity is set at $n = 3.23$, to be consistent with the effective index of the GaAs disks in our experiment. 
First, we numerically investigated the resonant modes at $2 < kR < 5$ using the finite element method (COMSOL Multiphysics 3.5a).  
Only transverse-electric (TE) polarized modes are considered, because experimentally the lasing modes are typically TE polarized due to the stronger vertical confinement (index guiding) and larger amplification by the InAs quantum dots (QDs) in the GaAs disks.          
In Fig. 1, the squares represent the highest $Q$ modes within the frequency range of interest in a cavity of $\epsilon = 0.26$, and the circles represent the lower $Q$ modes. 
The former (HQMs) resemble whispering gallery modes of radial number equal to one, whereas the latter (LQMs) have radial number equal to two.  
The $Q$ values for both modes decrease with decreasing $kR$. 
For $3.6 < kR < 4.1$, the HQMs nearly coincide with the LQMs in frequency, and their coupling results in a dip in the $Q$ of the HQM series. 
Away from the coupling regime, the HQMs have better output directionality than the LQMs, as characterized by $U \equiv \int I(\theta) \cos \theta d \theta / \int I(\theta) d \theta$, where $I(\theta)$ represents the angular distribution of the far-field intensity.  
[$U=0$ corresponds to isotropic or bidirectional emission, whereas positive (negative) $U$ corresponds to emission primarily towards $\theta=0^\circ\; (180^\circ)$.]   
This behavior is the opposite of what was observed previously in the wave regime of larger $kR$ (e.g., $5< kR < 25$), where the HQMs have lower $U$ than the LQMs except when they are coupled \cite{song_PRL10}.
In the semiclassical regime ($kR \gg 1$), the HQMs in the lima\c{c}on cavity have unidirectional emission, which is dictated by the chaotic ray dynamics in an open cavity \cite{wiersig_PRL08}. 
The universal directionality diminishes at smaller $kR$ when the ray model fails, and only accidental coupling to a directional LQM makes the output from a HQM unidirectional at $5< kR < 25$ \cite{song_PRL10}. 
The results in Fig. 1 illustrate that deeper into the wave regime ($ 2 < kR < 5$) the HQMs regain directionality when they are {\it decoupled} from the LQMs.
In fact, coupling to the LQMs is detrimental, because the LQMs are no longer directional.
Such behavior is desirable as it means that the lasing modes, which usually correspond to the HQMs, can have directional emission without the $Q$ reduction (and increased lasing threshold) associated with coupling to the LQMs. 

\begin{figure}[htbp]
\includegraphics[width=0.5\textwidth]{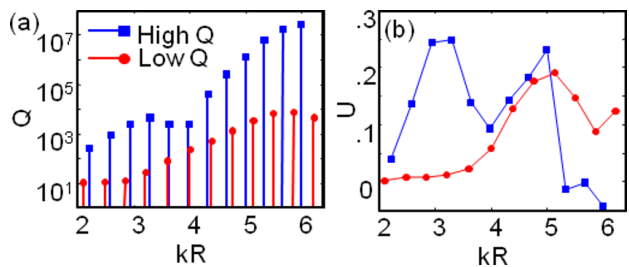}
\caption{(color online). Calculated $Q$ factor (a) and emission directionality $U$ (b) of the highest $Q$ modes (squares) and lower $Q$ modes (circles) as a function of $kR$ in a lima\c{c}on cavity of $\epsilon = 0.26$ and $n= 3.23$. The highest $Q$ modes have better emission directionality than the lower $Q$ modes deep in the wave regime $2 < kR < 5$, except when the two are coupled at $3.6< kR < 4.1$. }
\end{figure}

To understand the behavior of the HQMs, we analyzed their spatial profile and Husimi distribution.
The Husimi distribution projected onto the classical surface of section (SOS) provides the ray content of a mode at the cavity boundary in terms of the density of rays and their angle of incidence \cite{tureci_ProgressInOptics05, hentschel_EPL03}. 
Figure 2 presents the result for a HQM at $kR = 3.3$. 
The Husimi map indicates more clearly than the real-space intensity distribution that this mode has enhanced density near the three bouncing points on the cavity boundary of a ``triangle'' orbit.  
It is surprising to see the underlying ray orbit for a mode of such small $kR$. 
The maxima in the Husimi distribution, however, deviate from the exact locations of the bouncing points predicted by the ray orbit (marked by crosses) in the SOS.
This deviation originates from the openness of the system, which distinguishes incident and emergent (reflected) rays. 
A qualitative explanation can be given by two wave effects: ``Goos-H\"{a}nchen'' shift (GHS)  \cite{schomerus_PRL06,wiersig_PRE08, altmann_EPL08, wiersig_PRE10}, a lateral displacement of a beam total-internally reflected from a dielectric interface; and ``Fresnel Filtering'' (FF), a deflection of the reflected beam away from the specular reflection direction due to the angular spread of the incident beam \cite{rex_PRL02,tureci_OL02}.  
The GHS manifests as a horizontal shift in the SOS between the maxima of the Husimi projections for the incident and emergent rays at one bouncing point, whereas the FF induces a vertical shift.
Figure 2(c,d) only show the upper half ($\sin \chi >0$) portion of the SOS, which corresponds to CW circulating rays; the Husimi function in the lower half ($\sin \chi <0$, CCW rays) is reflected around $\theta=180^{\circ}$. 
From the locations of intensity maxima in the incident and emergent Husimis, we extract the CW and CCW orbits and plot them in Fig. 2(b).
The triangle orbit breaks into distinct CW and CCW periodic orbits, as shown in Fig. 2(b). 
The CW (CCW) orbit has a larger angle of incidence at bounce~3 (1) [labeled in Fig. 2(b)] than does the CCW (CW) orbit, leading to unidirectional emission towards $\theta = 0^\circ$ dominated by the CCW (CW) beam.

We also studied other HQMs, e.g. the one at $kR = 4.6$ which is based on an opposite triangle orbit
[approximate reflection of the triangle orbit in Fig. 2(b) through a vertical axis].  
Naively, we would expect the emission direction to switch from $\theta = 0^\circ$ to  $\theta = 180^\circ$. 
However, this is not the case, and the emission remains in the $\theta = 0^\circ$ direction. 
Since the lima\c{c}on cavity does not have reflection symmetry with respect to any vertical axis, the CW and CCW orbits must adjust themselves in the presence of GHS and FF to close after one round trip.
The emission directionality can be explained by the difference in the incident angles for the CW and CCW waves on the cavity boundary. 
Note that the underlying orbit for the mode at $kR = 3.3$ is a stable periodic orbit (SPO), while the one at $kR = 4.6$ is an unstable periodic orbit (UPO). 
In the semiclassical regime, the physical mechanisms for light escape from modes based on SPO and UPO are different: the UPO is determined by chaotic diffusion along the unstable manifolds, and the SPO by direct tunneling or chaos-assisted tunneling. 
In contrast, deep in the wave regime, the stability of the underlying ray orbits no longer plays a role in the output directionality, which instead is dominated by the wave effects (GHS and FF).   
We varied the deformation $\epsilon$ of the lima\c{c}on cavity up to 0.41 and observed similar behavior for the HQMs in spite of a dramatic change in the ray dynamics. 

\begin{figure}[htbp]
\includegraphics[width=0.5\textwidth]{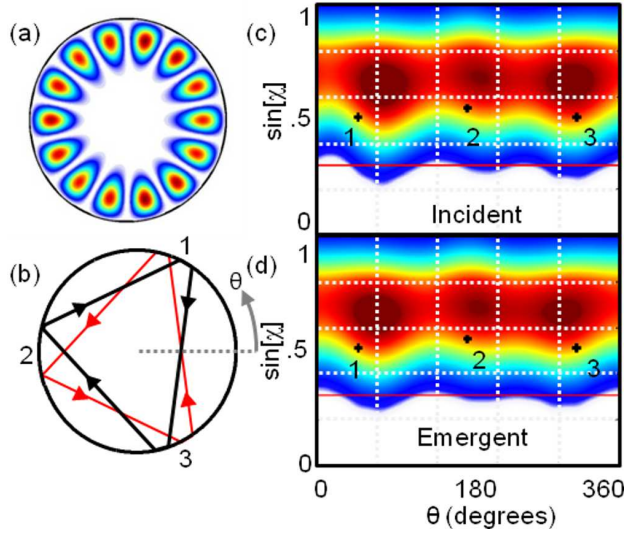}
\caption{(color online). Calculated spatial intensity distribution (a) and Husimi projections of the incident (c) and emergent/reflected (d) rays in the SOS for a high $Q$ mode with $kR = 3.3$ in the same lima\c{c}on cavity as Fig. 1. The horizontal axis in (c,d) is the polar angle $\theta$ that specifies the location of incident and reflected rays at the cavity boundary, the vertical axis is $\sin \chi$, where $\chi$ is the angle of incidence or reflection of the rays. The total internal reflection angle is indicated by the solid red line.  In both real space and SOS, this mode displays enhanced intensities near the three bouncing points [marked by crosses in (c,d)] of a triangle orbit. However, the CW and CCW propagating ray orbits (b), extracted from the locations of intensity maxima in the incident and emergent Husimi maps, are spatially separated due to the GHS and FF effects.}
\end{figure}

Next, we experimentally confirm the unidirectional emission from the lasing modes in ultrasmall lima\c{c}on cavities. 
Our sample fabrication procedure and lasing experimental setup are similar to those in ref. \cite{song_PRA09}.
Briefly, GaAs disks of lima\c{c}on shape were fabricated by electron-beam lithography, reactive ion etching and selective wet chemical etching.
The disks are 265 nm thick and supported by Al$_{0.7}$Ga$_{0.3}$As pedestals in the center.
InAs QDs embedded in the GaAs disk were optically excited to provide gain for lasing. 
Figure 3 presents the lasing result for a lima\c{c}on cavity of $R = 460$ nm and $\epsilon = 0.41$. 
A scanning electron microscope (SEM) image of this disk (top view) is shown in the inset of Fig. 3(a).  
The emission peak in the spectra of Fig. 3(a) displays a threshold behavior in its intensity as a function of the pump power [Fig. 3(b)]. 
The measured far-field pattern [inset of Fig. 3(b)] reveals that the emission is predominantly in the $\theta = 0^{\circ}$ direction, in good agreement with the numerical simulation. 
In addition to this mode with $kR = 3.3$, another mode at $kR = 2.98$ ($\lambda = 970 nm$) also exhibited unidirectional emission towards $\theta = 0^{\circ}$ (not shown).
We repeated the lasing experiments with lima\c{c}con cavities of different sizes, and observed similar phenomena deep in the wave regime.     

\begin{figure}[htbp]
\includegraphics[width=0.5\textwidth]{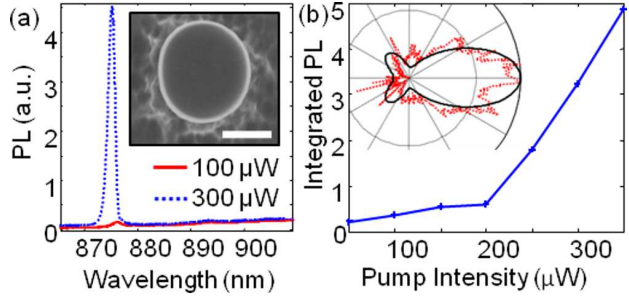}
\caption{(color online). (a) Measured photoluminescence (PL) spectra from a lima\c{c}on cavity of $R = 460$ nm and $\epsilon = 0.41$ at two incident pump powers: 100 $\mu$W (red solid line) and 300 $\mu$W (blue dotted line). The inset is the top-view SEM image of the disk, and the scale bar is 500 nm. (b) Intensity of the emission peak at $\lambda = 875$ nm in (a) as a function of the incident pump power. The threshold behavior indicates the onset of lasing action. The inset is a polar plot of the measured far-field pattern of laser emission (red dotted line) which agrees well with the calculated output of the high $Q$ mode at the same wavelength (black solid line).  }
\end{figure}

The spatial separation of CW and CCW waves in ultrasmall deformed cavities also offers an opportunity to selectively couple light from just one of them to a waveguide placed tangentially to the cavity boundary, providing dominant output in {\it one} waveguide direction. 
To optimize the position of the waveguide, we numerically studied the dependence of evanescent coupling on the waveguide location along the boundary. 
Figure 4 presents the results for a cavity whose shape is close to the lima\c{c}on and described in the polar coordinates as $r(\theta) = R [1 + \epsilon \cos(\theta)] [1 - \epsilon_1 \cos(2 \theta)] + d$, where $R$ = 890 nm, $\epsilon = 0.28$, $\epsilon_1 = 0.06$, and $d = 60$ nm. 
A straight waveguide with the same refractive index as the disk is separated from the disk boundary by 100 nm and the location of the coupling point is specified by the polar angle $\theta$.  
At each $\theta$, we calculated the steady-state intensity of emission coupled to the waveguide in the CW and CCW directions. We present the behavior of a mode with $kR=4.1$ although similar behavior was observed for nearby modes.
We found that directional coupling is possible provided the waveguide is positioned in a region without local chiral symmetry.
For example, if the waveguide is positioned at $\theta = 45^{\circ}$, where the local amplitude of the CCW orbit is highest, emission is mostly coupled in the CCW direction [Fig. 4(a)]. 
Figure 4(b) plots the directionality of the coupled emission $V$ as a function of the coupling position $\theta$. 
$V$ is defined as $(I_{CCW} - I_{CW}) / (I_{CCW} + I_{CW})$, where $I_{CCW}$ ($ I_{CW}$) represents the intensity of output coupled to the waveguide in the CCW (CW) direction. 
As the coupling point moves along the cavity boundary, the sign of $V$ changes, indicating a switch in the direction of the evanescent waveguide coupling.   
To confirm that the directional coupling results from local chirality, we calculated the intensities of the CW and CCW waves at each position (denoted by $\theta$) on the boundary by integrating the husimi intensities over all angles of incidence. Unlike refractive escape, evanescent coupling occurs for rays incident at angles well above the TIR angle, we therefore integrated over all angles of incidence to compare the total intensity of a given orbit available for evanescent coupling. As shown in Fig. 4(d), the CW and CCW intensities inside the cavity (without a waveguide) vary along the cavity boundary, and are shifted relative to each other, introducing a local unbalance between the CW and CCW waves.
Hence, locally there is a dominant energy flow in either the CW or CCW direction, mirroring the directional output to the waveguide in Fig. 4(b).      

\begin{figure}[htbp]
\includegraphics[width=0.5\textwidth]{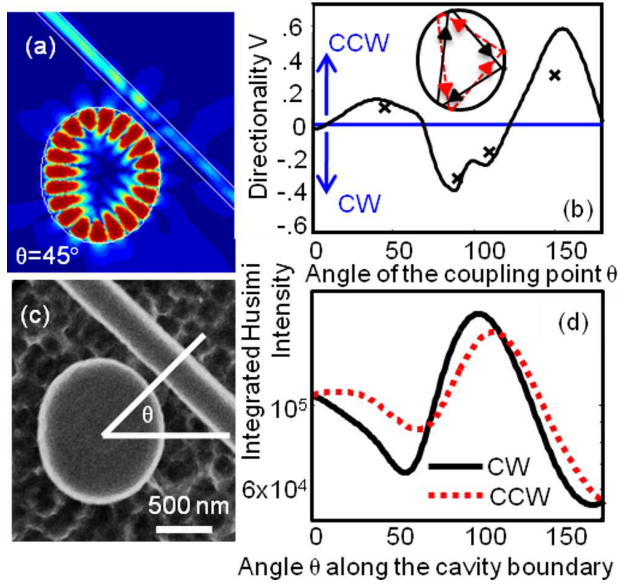}
\caption{ (color online). (a) Calculated intensity distribution in a waveguide coupled deformed microdisk, showing directional coupling to a waveguide positioned at polar angle $\theta = 45^{\circ}$.  (b)  Directionality of waveguide coupling $V$ as a function of the coupling position $\theta$ on the cavity boundary. $V > 0$ ($V < 0$) corresponds to stronger coupling in the CCW (CW) direction. The crosses represent the experimental data points which agree well with the numerical simulation (solid line). The inset shows the CW and CCW ray orbits for this cavity. (c) Top-view SEM image of a GaAs disk coupled to a waveguide. The disk is supported on an Al$_{0.7}$Ga$_{0.3}$As pedestal at the center, and the GaAs waveguide is free standing in air and supported by two Al$_{0.7}$Ga$_{0.3}$As pedestals at the ends. The background shows the residual left on the GaAs substrate after selective etching of the Al$_{0.7}$Ga$_{0.3}$As. (d) Husimi intensity integrated over $\sin \chi$ for the CW (black solid line) and CCW (red dotted line) waves as a function of location on the cavity boundary denoted by $\theta$. Locally, the CW and CCW intensities are not equal, leading to directional output to the waveguide shown in (b). Only half of the cavity boundary is plotted in (b,d), the other half can be obtained by mirror symmetry.}
\end{figure}

To confirm this behavior experimentally, we fabricated waveguide-coupled cavities using the same procedure described above. 
A 250 nm wide GaAs waveguide was separated from the disk edge by a 100 nm air gap, and the coupling positions were chosen to yield directional coupling. 
Figure 4(c) shows a SEM image of one of the GaAs disks with a coupled waveguide at $\theta = 45^{\circ}$. 
The cavity shape is identical to the simulated one in Fig. 4(a).
The waveguide was suspended in air by two Al$_{0.7}$Ga$_{0.3}$As pedestals (not shown). 
We achieved lasing with optical pumping and observed similar threshold behavior to what was presented above for the cavity without a waveguide. 
To monitor the directional coupling, we imaged the emission scattered by the two pedestals at the ends of the waveguide.  
A narrowband interference filter was placed in front of the camera to select a single lasing mode for imaging.  
By comparing the intensities of scattered light from the two pedestals, we calculated $V$.  
The experimental data points for the same cavity mode coupled to the waveguide at four different locations are plotted in Fig. 4(b) and are in good agreement with the numerical simulations. 
We characterized the waveguide coupling of several cavities of different sizes, and obtained similar directional coupling. 

Finally, we emphasize that the geometric shape of the cavities studied above maintains the chiral symmetry; that is, the cavity shape remains the same when $\theta$ is changed to $- \theta$. 
Therefore, the cavity resonances preserve the {\it global} chiral symmetry, i.e., the total intensities of the CW and CCW waves are equal.
This is confirmed for the cavity in Fig. 4(d) by integrating the CW or CCW intensities along the cavity boundary (from $\theta=0^{\circ}$ to $360^{\circ}$). 
Hence, chirality is introduced only locally via the unbalanced amplitudes of local CW and CCW waves along the boundary [Fig. 4(d)]. 
This allows us to selectively couple either the CW or CCW wave by adjusting the waveguide location on the cavity boundary.

In conclusion, we have achieved lasing in deformed cavities of $kR$ as small as $3$.
Even at such small values of $kR$, we are able to identify the ray trajectories underlying the resonant modes, and demonstrate that the CW and CCW propagating periodic orbits are spatially separated by the Goos-H\"{a}nchen shift and the Fresnel filtering effect.
The breaking of the local chiral symmetry between the CW and CCW waves not only leads to unidirectional emission in free space, but also makes the evanescent coupling to a waveguide directional.  
These results illustrate that local chirality can be created by wave optics effects in ultrasmall resonators and utilized to control the output directionality and enhance the emission collection efficiency. 

We thank Profs. A. Douglas Stone and Eugene Bogomolny for stimulating discussions. 
This work is supported partly by NIST under the Grant No. 70NANB6H6162, by NSF under the Grant Nos. ECCS-1068642 and ECCS-1128542, and by the DFG research group 760.

\bibliography{small_limacon_v6}

\end{document}